\documentclass{article}
\usepackage{amssymb}
\usepackage{epsfig}
\usepackage{subeqnarray}
\newcommand\DD {{\cal D}}
\def\res{\mathop{res}\limits_}

\renewcommand{\theequation}{\arabic{section}-\arabic{equation}}
\newtheorem{theorem}{Theorem}[section]
\newtheorem{remark}{Remark}[section]
\newtheorem{proposition}{Proposition}[section]
\newcommand{\eq}[1]{eq.~(\ref{#1})}
\newtheorem{definition}{Definition}[section]
\def\le{\left}
\def\m{\mathop}
\def\ri{\right}
\def\br{\begin{remark}\rm\small}
\def\1{{\bf 1}}
\def\er{\end{remark}}
\def\bt{\begin{theorem}}
\def\et{\end{theorem}}
\def\bd{\begin{definition}}
\def\ed{\end{definition}}
\def\bp{\begin{proposition}\rm}
\def\ep{\end{proposition}}
\def\be{\begin{equation}}
\def\ee{\end{equation}}
\def\bea{\begin{subeqnarray}}
\def\eea{\end{subeqnarray}}
\def \pa{\partial}
\def\C{{\mathbb C}}

\def\N{{\mathbb N}}

\begin{document}

\begin{flushright}
CRM-2841 (2002)\\
\hfill Saclay-T02/050\\
nlin.SI/0204054
\end{flushright}
\vspace{0.2cm}
\begin{center}
\begin{Large}
\textbf{Partition functions for Matrix Models and \\[10pt] Isomonodromic Tau
functions}\footnote{
Work supported in part by the Natural Sciences and Engineering Research Council
of Canada (NSERC) and the Fonds FCAR du Qu\'ebec.}
\end{Large}\\
\vspace{1.0cm}
\begin{large} {M.
Bertola}$^{\dagger\ddagger}$\footnote{bertola@crm.umontreal.ca}, 
 { B. Eynard}$^{\dagger
\star}$\footnote{eynard@spht.saclay.cea.fr}
 and {J. Harnad}$^{\dagger \ddagger}$\footnote{harnad@crm.umontreal.ca}
\end{large}
\\
\bigskip
\begin{small}
$^{\dagger}$ {\em Centre de recherches math\'ematiques,
Universit\'e de Montr\'eal\\ C.~P.~6128, succ. centre ville, Montr\'eal,
Qu\'ebec, Canada H3C 3J7} \\
\smallskip
$^{\ddagger}$ {\em Department of Mathematics and
Statistics, Concordia University\\ 7141 Sherbrooke W., Montr\'eal, Qu\'ebec,
Canada H4B 1R6} \\ 
\smallskip
$^{\star}$ {\em Service de Physique Th\'eorique, CEA/Saclay \\ Orme des
Merisiers F-91191 Gif-sur-Yvette Cedex, FRANCE } \\
\end{small}
\bigskip
\bigskip
{\bf Abstract}
\end{center}
For one-matrix models with polynomial potentials, the explicit relationship
between the partition function and the isomonodromic tau function for the
$2\times 2$ polynomial differential systems satisfied by the associated
orthogonal polynomials is derived.

\section{Introduction}
It is well-known \cite{orlov, morozov, vM} that  the partition
functions for matrix models, both for finite $n\times n$ random  matrices and
suitable $n\to \infty$ limits,  provide $\tau$ functions for the KP hierarchy.
It is also known that such partition functions satisfy certain constraints
amounting to the invariance of the  corresponding elements in the associated
infinite Grassmannian under a set of positive Virasoro generators.
This generally implies that there exist differential operators in the
spectral parameter variable which annihilate the associated
Baker-Akhiezer functions. The compatibility of the differential
equations in the spectral parameter and in the deformation parameters
representing KP-flows implies that the monodromy of the
operator in the spectral parameter is invariant under such
deformations. In the large $n$ limit this was considered in
\cite{moore}, while for finite $n$ it was studied in \cite{stringIts}.
Associated to  the analyticity properties of the Baker--Akhiezer
functions is an isomonodromic tau function \cite{JMU}, defined similarly to
the KP tau function. 

   In practice, the exact relationship between such ``isomonodromic'' tau
functions and the KP tau function is not very clearly understood. There are
special cases, however (see e.g. \cite{ASvM, AvM, aratyn, FIK, H1, HI,
isomIts, stringIts, JMU, moore}) where these two quantities can either be shown
to be identical, or at least explicitly related. In the computation of spacing
distributions for random matrices, a class of isomonodromic tau
functions occurs, given by Fredholm determinants of ``integrable''
integral operators \cite{BD, HI, HTW,  palmer, TW1, TW2} supported on a
union of intervals, with the deformation parameters taken as the endpoints of
the intervals rather than as KP flow parameters. The precise relation between
such deformations and the  KP flows is not yet clearly understood. (See however,
\cite{ASvM, AvM,  H1, vM} for some indications of the links between the two.)  

It is our purpose here to analyze in detail, for a generalization of
the simplest version of the Hermitian $1$-matrix model, with measures that
are exponentials of polynomial trace invariants, exactly what this
relationship is. The main idea is to consider, for finite $n$ (the size of the
random matrix), the isomonodromic deformation systems satisfied by the
associated sequence of orthogonal polynomials, and to compute an explicit
formula (Thm. \ref{mainthm}) relating the associated isomonodromic tau function
to the partition function of the generalized matrix model. In the process, we
recall (and generalize) some of the  standard results concerning the relation
of isomonodromic deformations to random matrices and the occurrence of Stokes
phenomena in solutions to  the associated systems of differential equations
with polynomial coefficients.

\section{Orthogonal polynomials for semiclassical functionals}
We consider the orthogonal polynomials for a moment functional 
\bea
&& \mathcal L_{\varkappa\Gamma}: \C[x]\mapsto\C\ , \\
&& \mathcal L_{\varkappa\Gamma}(x^i)
:= \int_{\varkappa\Gamma} {\rm d}x {\rm e}^{-V(x)} x^i\ ,
\label{functional}
\eea
where the integration over $\varkappa\Gamma$ represents a suitable linear
combination of integrals over contours to be explained in detail below.
The function $V(x)$ is called the {\em potential} and it will be taken here 
to be a polynomial of degree $d+1$,
\be
V(x):= \sum_{K=1}^{d+1} u_K\frac {x^K}{K}\ ,
\ee
in which the coefficients $\{u_K\}$ are viewed as deformation parameters.
(The constant term is suppressed since it only contributes to the
overall normalization.)

This is a generalized setting which reduces to the ordinary orthogonal
polynomials when the measure is positive and the contour of integration is on 
the real axis.  We do not assume that $V(x)$ is of even degree or that it
is  real; the moment functional (\ref{functional}) is in general complex-valued.
In this general setting the orthogonal polynomials (if
they exist) are defined as follows.
\bd
\label{OPs}
The (monic) orthogonal polynomials $\{p_n(x)\}_{n\in\N}$ are defined by
the requirements
\bea
\mathcal L_{\varkappa\Gamma}(p_n(x)p_m(x))\!\!\! &\!=\!&\!\!\! h_n\delta_{nm}\ , \ \ h_n\in
\C^\times\slabel{11}\\
p_n(x)\!\!\!&\!\!\!\!=\!\!\!\!&\!\!\! x^n+\dots\ .\slabel{12}
\eea
\ed
It is easily seen \cite{Szego} that such polynomials exist if and only if the
Hankel determinants formed from  the moments do not vanish
\be
\Delta_n(\varkappa\Gamma,V):= \det\le(\mathcal L_\Gamma(x^{i+j})\ri)_{0\leq
i,j\leq n-1} \neq 0\ , \ \ \forall n\,\in\, \N\ .
\ee 
(If they do exist, eqs. (\ref{11}, \ref{12}) determine them uniquely.)

 We now return to the definition of the cycle $\varkappa \Gamma$. The convergence
of the integral defining the moment functional $\mathcal L$ implies
that the contour of integration  must extend to infinity in such a way that
the real part of $V(x)$ diverges to $+\infty$. For the purpose of the
description of $\varkappa\Gamma$
let us introduce the sectors
\begin{figure}
\epsfxsize 12cm
\epsfysize 9cm
\centerline{\epsffile{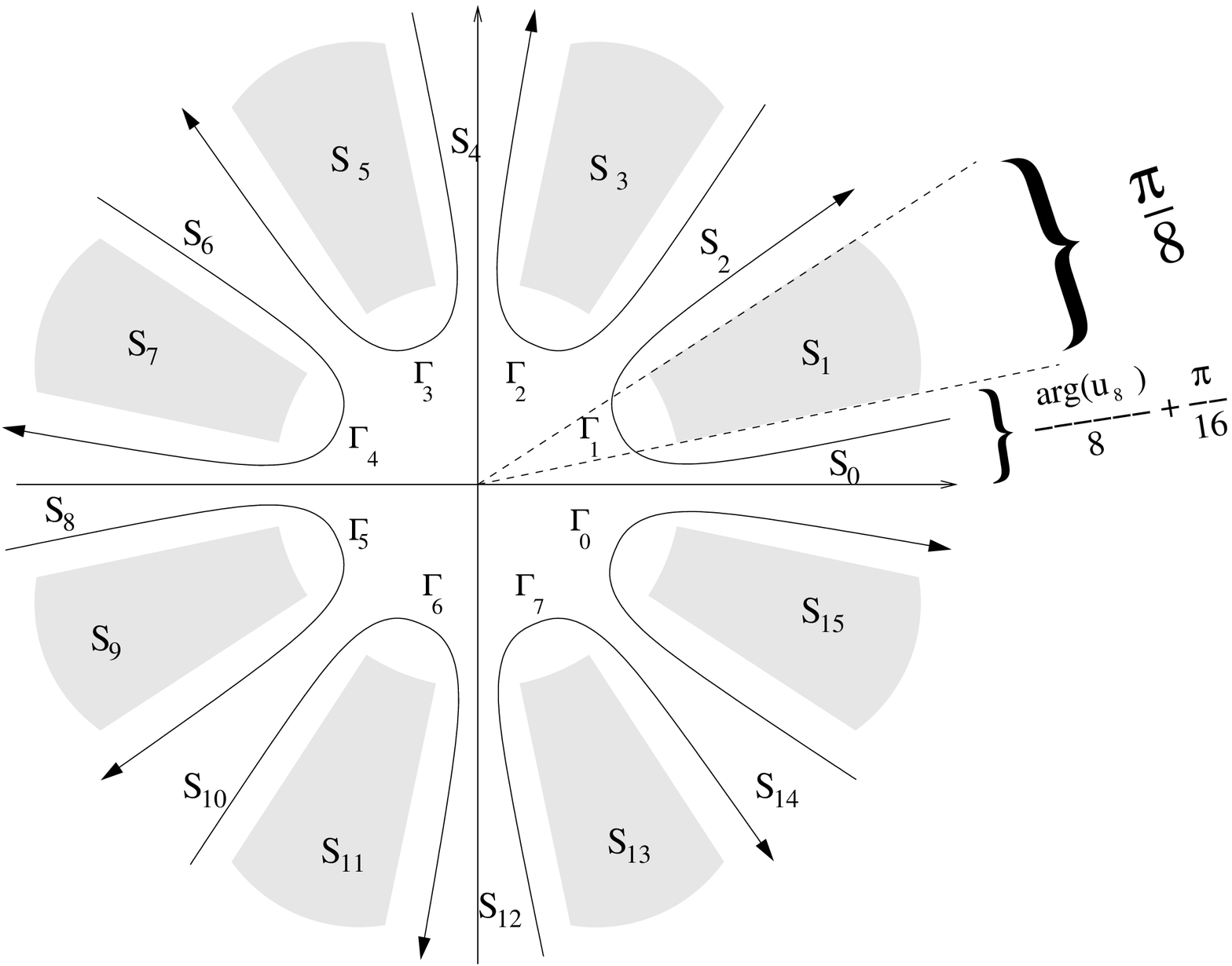}}
\caption{An example of the contours $\Gamma_k$ in the case $d=7$.}
\label{figone}
\end{figure}
\begin{eqnarray}
\mathcal S_k:=\le\{x\in \C\ ,\ {\rm arg}(x)\in
\le(\vartheta+\frac{(2k-1)\pi}{2(d+1)}, \vartheta+\frac{(2k+1)\pi}{2(d+1)} \ri)\ri\},\
\label{sect}\\
\vartheta:= \frac{\arg(u_{d+1})}{d+1}\ ,\ \ k=0,\dots, 2d+1\
.\nonumber
\end{eqnarray}
These sectors are defined in such a way that $\Re(V(x))\to+\infty$ in
the even numbered sectors. Therefore a contour should come from infinity
in an even sector and return to infinity in a different even sector.
 Such functionals have been studied in a similar setting in \cite{marce,
marce1}.

We define the contours $\Gamma_k, \ k\in {\bf Z}_d$ as the ones coming from
infinity in the sector $\mathcal S_{2k-2}$ and returning to infinity in
the sector $\mathcal S_{2k}$ (see Figure \ref{figone}). It  follows from the Cauchy 
theorem that the corresponding $d+1$ functionals $\mathcal L_{\Gamma_k}$ satisfy
\be
\sum_{k=0}^{d}\mathcal L_{\Gamma_k} = 0\ ,
\ee
but it can be shown that any $d$ of them  are linearly
independent \cite{marce}.
 In general we may consider any linear combination of such
moment functionals corresponding to a ``one-cycle'' and the
corresponding deformation theory under changes of complex generalized
measure. The quantities $\varkappa\Gamma$ and $\mathcal L_{\varkappa\Gamma}$ 
are understood to mean: 
\bea
&& \varkappa\Gamma:=
\sum_{k=1}^d \varkappa_k \Gamma_k\ ,\label{kappagamma}\\
&& \mathcal L_{\varkappa\Gamma}(\cdot):= \sum_{k=1}^d
\varkappa_k\mathcal L_{\Gamma_k} (\cdot):= 
\sum_{k=1}^d \varkappa_k \int_{\Gamma_k}\!\!\!\!{\rm d}x \, {\rm
e}^{-V(x)}(\cdot)=:\int_{\varkappa\Gamma}\!\!\!\!{\rm d}x\, {\rm
e}^{-V(x)}(\cdot)\ ,
\eea
where $\{\varkappa_k\in \C\}_{k=1,\dots d}$ are any fixed set of constants (not
all vanishing). (A similar setting, in which the contours $\Gamma_k$ play the
roles of boundaries between Stokes sectors, was considered, e.g.,
in the earlier works  \cite{BlIt, FIK, isomIts, stringIts}.

 For a fixed potential $V(x)$ the Hankel determinants
$\Delta_n(\varkappa\Gamma,V)$  are homogeneous polynomials in
$\varkappa$ of degree $n$. The nonvanishing of this denumerable set of 
polynomials determines a set with full (Lebesgue) measure in the $\varkappa$
space and in this sense the orthogonal polynomials exist for generic
$\varkappa$. (An interesting problem, which  will not be addressed here, is
whether the set of $\varkappa$'s for which the orthogonal polynomials exist
contains  an open set, thus allowing for arbitrary infinitesimal deformations.)

If the contour of integration is not the real axis this setting does not derive
in general from a matrix model. Nevertheless,  we introduce what we call the
{\em partition function}, in analogy with the Hermitian matrix model case
\cite{mehta, Szego},
 through the following formula
\be
\mathcal Z_n:= \int_{\varkappa \Gamma}\!\!\!\!\!{\rm
d}x_1\cdots\int_{\varkappa \Gamma}\!\!\!\!\!{\rm d}x_n 
\, \Delta(\underline x)^2 {\rm e}^{-\sum_{j=1}^n V(x_j)} =
n!\Delta_n(\varkappa,V)\  ,\ \ \Delta(\underline x):= \prod_{i<j}
(x_i-x_j)\ .\label{partit}
\ee
It is well known that $\mathcal Z_n$ is proportional to the product
of all normalization factors for the monic orthogonal polynomials
in Def. \ref{OPs}
\be
\mathcal Z_n = n!\prod_{j=0}^{n-1} h_j\ .
\ee
Instead of the $p_n$'s, it is convenient to introduce the following
quasi-polynomial functions:
\be\label{defpsin}
\psi_n(x) = {1\over \sqrt{h_n}}\, p_n(x)\, {\rm e}^{-{1\over 2} V(x)} \ ,
\ee
which are orthonormal, in the sense that 
\be\label{ortho}
\int_{\varkappa\Gamma}\!\!\!\! {\rm d}x\, \psi_n(x) \psi_m(x) = \delta_{nm} \ .
\ee

  These (quasi-)polynomials satisfy a three-term recursion relation
\cite{Szego}: 
\be\label{multx}
x \psi_n(x) = \sum_{m=0}^\infty Q_{nm} \psi_m  = \gamma_{n+1}
\psi_{n+1} + \beta_n \psi_n + \gamma_n \psi_{n-1} \ , 
\ee
where 
\be\label{gammahn}
\gamma_n = Q_{n,n-1} = \sqrt{h_{n}\over h_{n-1}}\neq 0 \ \qquad \forall
n\in \N,\ {\rm and}\ \  \beta_n = Q_{nn}  \ .
\ee
The semi-infinite matrix $Q$ is symmetric ($Q_{nm}=Q_{mn}$), and has
its nonzero entries along a diagonal band of width $3$
 ($Q_{mn}=0$ if $|m-n|>1$); i.e., it is a symmetric Jacobi matrix:
\be\label{defQ}
Q = \pmatrix{\beta_0 & \gamma_1 & 0 & \dots & \dots \cr
             \gamma_1 & \beta_1 & \gamma_2 & 0 & \cr
             0 & \gamma_2 & \beta_2 & \gamma_3 & \cr
             \vdots & 0 & \ddots & \ddots & \ddots 
} \ .
\ee
These recursion relations may equivalently be expressed as a 
$2$-component vector recursion:
\be
\le[\matrix{\psi_{n}(x) \cr \psi_{n+1}(x)}\ri] = R_n(x)
\le[\matrix{\psi_{n-1}(x)\cr \psi_{n}(x)}\ri]\ ,\label{2receq}
\ee
where
\be
R_n(x):= \le[ 
\begin{array}{cc}
0&1\\
 -\frac {\gamma_n}{\gamma_{n+1}} &\frac {x-\beta_n}{\gamma_{n+1}}
\end{array}\ri] \label{2rec}.
\ee
\indent We can also 
express the  derivatives of the $\psi_n$'s  in the basis $\left\{
\psi_m \right\}_{m=0\dots  \infty}$:
\be\label{derivx}
\psi_n'(x) = \sum_{m=0}^\infty P_{nm} \psi_m(x) \ ,
\ee
where $P$ has non--zero entries in a band of width $2d+1$ centered on
the principal diagonal.
An integration by parts using   \eq{ortho} shows  that the semi-infinite matrix
 $P$ is antisymmetric. From 
\be
p_n' = n p_{n-1} +
\mathcal O(x^{n-2})\ ,
\ee
 one obtains:
\bea
P_{nm} + {1\over 2} \left(V'(Q)\right)_{nm} & = & 0 \qquad {\rm if} \,\, m\geq n  \\
P_{n,n-1} + {1\over 2} \left(V'(Q)\right)_{n,n-1} & = & {n\over \gamma_n} \ ,
\eea
and hence 
\be\label{PV}
P = -{1\over 2} \left( V'(Q)_+ - V'(Q)_- \right) \ ,
\ee
where for any semi-infinite matrix $A$, $A_+$ (resp. $A_-$), denotes
the upper (resp. lower) triangular part of $A$. 
 Eq. (\ref{PV}) also  implies the following equations, referred to
sometimes as  ``string equations'' \cite{ZJDFG, doug, doug1, courseynard,
stringIts}: 
\be\label{eqmvt}
[Q,P]=\1\ ,\ \ 
0  =  V'(Q)_{nn}  \ \  ,\ \ {n\over \gamma_n}   =
V'(Q)_{n,n-1}\label{eqVg}\ . 
\ee

\subsection{Differential system}
It follows from (\ref{PV}) that the sum in (\ref{derivx}) contains
only a finite number 
of terms and can be written:
\be\label{psinprimeeta}
\psi_n'(x) = -{1\over 2} \sum_{k=1}^{d} \le( V'(Q)_{n,n+k} \psi_{n+k}
- V'(Q)_{n,n-k}\psi_{n-k}\ri) 
\ \ .
\ee
Using \eq{multx} recursively  for any $0<k \leq d$ we  can write $\psi_{n+k}$ 
and  $\psi_{n-1-k}$ as  a unique linear combination of $\psi_n$
and $\psi_{n-1}$ with coefficients that are  polynomials in $x$  of degree
 $\leq k$.
Eq. \ref{psinprimeeta} can thus be rewritten as a $2\times 2$ differential system:
\be
{{\rm d}\over {\rm d}{x}} \le[\matrix{\psi_{n-1}(x) \cr
\psi_{n}(x)}\ri] =  \DD_n (x)
\le[\matrix{\psi_{n-1}(x) \cr \psi_{n}(x)}\ri]\ ,\qquad n\geq 1 \label{eqDD}\ ,
\ee
where $\DD_n(x)$ is a $2\times 2$ matrix with polynomial coefficients of degree
at most $d=\deg V'$.
In fact we can choose the sequence of $\DD_n$'s to
satisfy the recursive gauge transformation relations 
\be
\DD_{n+1} =R_n \DD_{n} {R_n}^{-1} + \pa_x R_n{R_n}^{-1}\ .\label{recDD}
\ee
(See also \cite{BEH} for similar statements for the case of 2-matrix models.)
 This relation determines the sequence of $\DD_n$'s uniquely
\cite{bauldry, bonan, FIK, ismail, isomIts, stringIts}.
It follows in particular that these transformations preserve the
(generalized) monodromy of the linear systems (\ref{eqDD}) for all $n$.

\begin{theorem}\label{thDeriv}
The matrix $\DD_n(x)$ is given by the formula:
\bea\slabel{thDD}
\DD_n(x) \!\!\!&=&\!\!\!  {1\over 2} V'(x)  \le[\matrix{1 & 0 \cr 0 & -1}\ri] +
\le[\matrix{ \left( {V'(Q)-V'(x)\over Q-x}\right)_{n-1,n-1} & \left(
{V'(Q)-V'(x)\over Q-x}\right)_{n-1,n} 
\cr \left( {V'(Q)-V'(x)\over Q-x}\right)_{n,n-1} &   \left(
{V'(Q)-V'(x)\over Q-x}\right)_{nn}  }\ri] \, \le[\matrix{ 0 & -\gamma_n \cr
\gamma_n & 0 }\ri]    \\
\!\!\!&=&\!\!\!    {1\over 2} V'(x)  \le[\matrix{1 & 0 \cr 0 & -1} \ri]+
\gamma_n 
 \le[\matrix{\left(
{V'(Q)-V'(x)\over Q-x}\right)_{n-1,n}  &-  \left( {V'(Q)-V'(x)\over
Q-x}\right)_{n-1,n-1}\cr
\left({V'(Q)-V'(x)\over Q-x}\right)_{nn} &  
-\left( {V'(Q)-V'(x)\over Q-x}\right)_{n,n-1}}\ri]\slabel{formulafordd}
\eea
\end{theorem}
Proofs of this result may be found in \cite{bauldry, bonan, ismail} and
earlier in \cite{stringIts} for the case of even potentials. In the Appendix we
give an alternative proof which is based upon the deformation
formulas obtained in Theorem \ref{thdeform} below.

\br
The matrix $\gamma_n\le[\matrix{0&-1\cr1&0}\ri]$ appearing in the first
line of \eq{formulafordd} arises naturally in the
Christoffel--Darboux formula written here for the quasi--polynomials
\be
(x-x')\sum_{j=0}^{n-1} \psi_{j}(x)\psi_{j}(x') =
\gamma_n\big(\psi_n(x)\psi_{n-1}(x')- \psi_{n-1}(x)\psi_n(x')\big) =
\gamma_n \le[\psi_{n-1}(x),\psi_{n}(x)\ri]\le[\matrix{0&-1\cr1&0}\ri]
\le[ \matrix{\psi_{n-1}(x')\cr \psi_n(x')}\ri] \ .
\ee
\er

Equation (\ref{recDD}) is the starting point that 
 allows us to explicitly compute the spectral curve of
the matrix  $\DD_n(x)$ in terms of the deformation parameters
$\{u_K\}_{K=1\dots d}$ and the partition function $\mathcal Z_n$.

In addition to \eq{eqDD} we may consider the effect of deformations of
the $\{u_K\}$ parameters on the orthogonal polynomials. It is
well-known that these give rise to isospectral deformations of the Jacobi
matrix $Q$ (Toda hierarchy equations \cite{orlov, morozov})
\bea
&& \frac \pa{\pa u_K} Q = \le[U_K,Q\ri]\ ,\slabel{definfty0}\\ 
&& \frac \pa{\pa u_K}\psi_n(x) = \sum_{m} (U_K)_{nm}\psi_m(x)
\slabel{definfty1}\ ,\\ 
&& U_K:= -\frac 1 K
\le({Q^K}_+- {Q^K}_-\ri)\ .\slabel{definfty}
\eea
Combining the recursion relations (\ref{multx}) with  equations
(\ref{definfty0}, \ref{definfty1}, \ref{definfty}) one
obtains deformation equations for the successive pairs of orthogonal quasi-polynomials 
\be
\frac \pa{\pa u_K}\le[\matrix{\psi_{n}(x)\cr \psi_{n+1}(x)} \ri] =
\mathcal U_{K,n}(x) \le[\matrix{\psi_{n}(x)\cr \psi_{n+1}(x)} \ri]\
.\label{starstar} 
\ee
The matrices $\mathcal U_{K,n}(x)$ can be explicitly computed 
and chosen to have polynomial entries of degree $\leq K$ as given in
the following theorem, whose proof is given in the appendix. 
(Similar formulas for the case of even potentials were given in
\cite{stringIts}. The main simplification in that case is that the
coefficients $\beta_n$ vanish for all $n$'s.)
\begin{theorem} \label{thdeform}
The expression for ${\cal U}_{K,n}(x)$ may be taken as
\be
{\cal U}_{K,n}(x)  = {1\over 2K} \le[\matrix{ x^K-Q^K_{n-1,n-1}  & 0
\cr 0 & Q^K_{nn}-x^K }\ri] 
+{1\over K}\gamma_n \le[\matrix{    \left({x^{K}-Q^{K}\over
x-Q}\right)_{n,n-1}  &  - \left({x^{K}-Q^{K}\over
x-Q}\right)_{n-1,n-1}\cr \left({x^{K}-Q^{K}\over x-Q}\right)_{nn} &
-\left({x^{K}-Q^{K}\over x-Q}\right)_{n,n-1}  }\ri]\ .
\ee
\end{theorem}
\br
It follows as a corollary to Thm. \ref{thdeform} that:
\be
{\rm Tr}\, {\cal U}_{K,n}(x) = {\pa \over \pa u_K} \ln{\gamma_n} \ .
\ee
\er

 It is important to note that eqs. (\ref{2receq}), (\ref{eqDD}) and
(\ref{starstar}) do not just hold for the sequence $\{\psi_n\}$ of
orthogonal quasi-polynomials but define an overdetermined system of
differential-difference-deformation equations that are
 actually compatible in the Frobenius sense. Hence there exists,
for all $n$, a basis of simultaneous solutions of these equations. This 
system of solutions  can be written explicitly by adding to the $\psi_n$'s
a solution of the second type as follows
\bea
&& \Psi_n(x) = \le[\matrix{\psi_{n-1}(x) &  \widetilde \psi_{n-1}(x)\cr 
\psi_n(x) & \widetilde \psi_n(x)}\ri]\ ,\\
&& \widetilde \psi_n(x) = {\rm
e}^{\frac 12 V(x)}\int_{\varkappa\Gamma}\!\!\!\!{\rm d} x\,\frac { {\rm
e}^{-\frac 1 2 V(z)} \psi_n(z)}{x-z}\ .\slabel{fundsys}
\eea
It is an easy exercise to show that these $\widetilde \psi_n$'s satisfy the same
recursion relations as the $\psi_n$'s for $n>d$ and hence satisfy
the same ODE (\ref{eqDD}). It is also straightforward to verify that
they satisfy the deformation equations (\ref{starstar}).
The fact that these equations are compatible implies that the generalized
monodromy data (i.e. the Stokes matrices) for the operator $\pa_x - \DD_n(x)$ 
are independent of the deformation parameters $\{u_K\}$ and the integer $n$.

We now turn to the main object of this paper: to relate the
partition function (\ref{partit})  with the isomonodromic tau function (whose
definition is given below). The first step is to connect the
spectral curve of the matrix $\DD_n(x)$ with the partition function.
\bt
\label{spectral_partition}
The spectral curve is given by
\bea
0\!\!\!\!\! 
&&\!\!\!\!\!=\! \det(y\1\!- \!\DD_n(x))\! = \!y^2\!-\!\frac 1 2 {\rm Tr}
\le({\DD_n}^2(x)\ri) = y^2\!- \! \frac 14
(V'(x))^2 \!+\!\sum_{j=0}^{n-1} \le(\frac {V'(Q)\!-\!V'(x)}{Q-x}\ri)_{jj}\!\!\! \!\!\!  
\\
&&\!\!\!\!\!=\! y^2- \frac 14 (V'(x))^2 +n\sum_{J=1}^{d} u_{J+1}x^{J-1} +
\sum_{K=0}^{d-2} x^K \sum_{J=1}^{d-K-1}u_{J+K+2} J\frac \pa{\pa u_J}
\ln(\mathcal Z_n)\ .
\eea
\et
Before proving this result, we remark that the RHS contains matrix
elements of $Q$ with all indices down to $n=0$ while, as it stands,
formula (\ref{formulafordd}) for $\DD_n(x)$ contains only matrix
elements of $Q$ with indices 
``around'' $n$, i.e., at distances less than $d+1$ from $Q_{nn}$.\\
{\bf Proof of Thm. \ref{spectral_partition}}.
First, notice that $\DD_n(x)$ is traceless (by formula (\ref{formulafordd})).
Using eq. (\ref{recDD}) and the cyclicity of traces
of products one finds that
\be
 {\rm Tr}\le({\DD_{n+1}(x)}^2\ri) = {\rm Tr}\le({\DD_{n}(x)}^2\ri)
+2 {\rm Tr}\le(\DD_n(x){R_n}^{-1}(x)R_n'(x)\ri) +  {\rm
Tr}\le(\le(R_n'(x){R_n}^{-1}(x)\ri)^2\ri)\ . 
\ee
Now note that 
\be
R_n'(x){R_n}^{-1}(x) = \le[\matrix{ 0&0\cr \frac 1 \gamma_n &
0}\ri]\ ,\qquad\ {R_n}^{-1}(x)R_n'(x) = \le[\matrix{0&-\frac
1{\gamma_n} \cr 0&0}\ri]\ ,
\ee
so the third trace term vanishes and the second one  gives 
\be
 {\rm Tr}\le({\DD_{n+1}(x)}^2\ri) = {\rm Tr}\le({\DD_{n}(x)}^2\ri) - 2
\le(\frac {V'(Q)-V'(x)}{Q-x} \ri)_{nn}\ .
\ee
Therefore 
\be
{\rm Tr}\le({\DD_{n}(x)}^2\ri) ={\rm Tr}\le({\DD_{1}(x)}^2\ri) -2
\sum_{j=1}^{n-1}\le(\frac {V'(Q)-V'(x)}{Q-x} \ri)_{jj}\ . 
\ee 
Note that $\DD_0$ is not defined. 
We therefore need to compute the trace of the square of the first matrix
$\DD_1$. 
To this end we introduce the notation
\be
W(x):= \frac {V'(Q)-V'(x)}{Q-x}\ .
\ee
Now use the string equation  (\ref{eqmvt}) 
to compute
\bea
&& -V'(x)=\le[V'(Q)-V'(x)\ri]_{00} = \le[W(x)(Q-x)\ri]_{00} = (\beta_0-x)
W_{00}(x) + \gamma_1 W_{10}(x)\slabel{uno}\\
&& \frac 1{\gamma_1}= 
\le[V'(Q)-V'(x)\ri]_{10} = \le[W(x)(Q-x)\ri]_{10} =
\gamma_1W_{11}(x) + (\beta_0-x)W_{10}(x)\ .\slabel{due}
\eea
Multiplying  eq. (\ref{uno}) by $\gamma_1 W_{10}$, eq. (\ref{due}) by
$\gamma_1W_{00}$ and subtracting, we  obtain (since the diagonal elements of
$V'(Q)$ vanish  by eq. (\ref{eqmvt})) 
\be
W_{00}(x) =
{\gamma_1}^2W_{11}(x)W_{00}(x)-\gamma_1W_{10}(x)\bigg(V'(x)+\gamma_1
W_{10}(x)\bigg)\label{tre}\ .
\ee 
Computing the trace of the square of $\DD_1(x)$, and using
eq. (\ref{tre}) we obtain
\be
{\rm Tr}\,\le(\DD_1(x)^2\ri) =  \frac 1 2 V'(x)^2-2W_{00}(x)\ . 
\ee
Therefore the full formula for the trace of the square is 
\be
{\rm Tr}\le({\DD_{n}(x)}^2\ri) =\frac 1 2 {V'(x)}^2 -2
\sum_{j=0}^{n-1}\le(\frac {V'(Q)-V'(x)}{Q-x} \ri)_{jj}\ .
\ee
\br
Notice that, had we interpreted 
 formula (\ref{formulafordd}) for $n=0$ to mean that 
$\gamma_0=0$, we would have
obtained exactly the same result.
\er

We now expand the polynomial 
\be
\frac {V'(Q)-V'(x)}{Q-x} = \sum_{J=1}^{d} u_{J+1}x^{J-1}\1 +
\sum_{K=0}^{d-2} x^{K} \sum_{J=1}^{d-K-1}u_{J+K+2}Q^J
\ee 
and use the fact (proved in the appendix, eq. (\ref{traceUk})) that 
\be
\sum_{j=0}^{n-1} (Q^K)_{jj} = K\pa_{u_K} \ln(\mathcal Z_n)\ ,
\ee
to complete the proof of the theorem. Q.E.D.

\vskip 3pt
 
Theorem \ref{spectral_partition} relates the spectral curve of the
matrix $\DD_n(x)$ to the partition function $\mathcal Z_n$.
The latter  is known to be a $\tau$-function for the KP hierarchy.
(A short proof of this fact is given in Appendix \ref{B} together with
an interpretation of $\mathcal Z_n$ as a  Fredholm determinant.)
On the other hand, a linear system of the form (\ref{eqDD}) with
compatible deformation equations (\ref{2receq}), (\ref{starstar}),  
allows one to define an {\em isomonodromic tau function} \cite{JMU}.
We recall here briefly, in a way adapted to the case at
hand, how to define such a tau-function.

The ODE
\be
\frac {\rm d}{{\rm d}x} \Psi(x) = \DD_n(x)\Psi(x)\ \label{ODE}
\ee
has just one irregular singularity at $x=\infty$ and hence has  Stokes
sectors,
 which are just those defined in \eq{sect}.
 Since the leading term of the singularity  in $\DD_n(x)$ is
semisimple with distinct eigenvalues
\be
\DD_n(x) \simeq \frac {u_{d+1}}2 x^d\sigma_3 + \mathcal O(x^{d-1})\ ,
\ee
we can write in each Stokes sector $\mathcal S_k$  a formal
asymptotic solution of the form 
\bea
&& \Psi_n(x) \simeq C_n Y_n(x)\,{\rm e}^{T_n(x)} \ ,\\ 
&& Y_n(x) := \1 + \sum_{j=1}^\infty Y_{j,n}x^{-j} \ ,\ \\
&&T_n(x) = \sum_{K=1}^{d+1} \frac 1 K T_Kx^K +T_{-1}\ln(x)\ ,
\eea
where the $T_K$ are all diagonal and  (in our case) traceless. 
In the present case we easily find $T_n(x)$ from the tracelessness
and the exponential
part of the asymptotic form of the fundamental system defined in
\eq{fundsys} (cf. refs. \cite{BlIt, BlIt1} for similar formulas in
the case of quartic even potentials). That is, 
\bea
T_K = \frac 1 2 {\rm diag}(u_K,-u_K) =\frac { u_K}{2}\sigma_3,\ K>0,
\qquad T_{-1}=n\sigma_3\\
T_n(x) = -\le(\frac 1 2 V(x)-n\ln(x)\ri) \sigma_3\ .
\eea 
The Stokes matrices relate the basis of solutions with these asymptotic
forms in contiguous sectors.
The isomonodromic tau function is then defined by the
deformation equation. (See the introduction to \cite{JMU} for the general 
context and motivations behind the theory of isomonodromic tau functions.)
\be
{\rm d}\ln(\tau_{IM,n}) := 
\res{x=\infty} {\rm Tr} \le(Y_n^{-1}(x)Y_n'(x) {\rm d}T(x)\ri) = 
\sum_{K=1}^{d+1}\frac 1 {2K} \res{x=\infty} {\rm Tr}
\le(Y_n^{-1}(x)Y_n'(x)\,\sigma_3 
x^{K-1}\ri){\rm d}u_{K}
 \ ,\label{tauisom}
\ee
where
\be
\qquad {\rm d} := \sum_{K=1}^{d+1} {\rm d}u_K \frac
\pa{\pa u_K} \ .
\ee
We now state and  prove the main theorem relating $\tau_{IM,n}(\varkappa, V) $
and $\mathcal Z_n(\varkappa,V)$. 
\bt
\label{mainthm}
The partition function $\mathcal Z_n(V)$ is related to the isomonodromic tau
function by the formula 
\be
 {\tau_{IM,n}}(\varkappa,V)=\mathcal Z_n(\varkappa,V) {{u_{d+1}}^{\frac
{n^2}{d+1}}}\ , 
\ee
up to a multiplicative factor that  depends only on $\varkappa$. 
\et 
{\bf Proof}.
We will prove that 
$$ {\rm d}\ln\mathcal Z_n = {\rm d}\ln \tau_{IM,n} -
\frac{n^2}{(d+1)u_{d+1}}{\rm d}u_{d+1}$$
We start by observing that (suppressing the dependence on $x$ and $n$ for
brevity and denoting the derivative with respect to  $x$ by a prime)
\bea
\DD_n \!\!\!&=&\!\!\! \Psi'\Psi^{-1} = C\,Y'\,Y^{-1} C^{-1} + C \, Y\, T'\,
Y^{-1}C^{-1}\\
{\rm Tr}({\DD_n}^2)  \!\!\!&=&\!\!\! {\rm Tr}\le( (Y'Y^{-1})^2\ri) +{\rm Tr}\le(
{T'}^{2}\ri)   +2 {\rm Tr}\le(Y^{-1}Y'\, T' \ri)\ .
\eea
Consider the coefficient of $x^K$ in ${\rm Tr}({\DD_n}^2)$,
which is:
\begin{eqnarray}
&& \frac 1 2 \sum_{j=0}^{K} u_{j+1}u_{k+1-j} -2 n u_{K+2} -
2\!\!\!\!\!\sum_{J=1}^{d-K-1}u_{J+K+2} J\frac \pa{\pa u_J}
\ln(\mathcal Z_n) \m{\hbox{\huge =}}^{\hbox{\scriptsize Thm. \ref{spectral_partition}}}  
-\res{x=\infty} x^{-K-1}{\rm Tr}({\DD_n}^2) =\cr
&&=\overbrace{ -\res{x=\infty} x^{-K-1} {\rm Tr} \le(
(Y'Y^{-1})^2\ri)}^{=0 \hbox{ \scriptsize  because } Y'Y^{-1} = \mathcal O(x^{-2})}  - 2\res{x=\infty} x^{-K-1} 
\le(\frac 1 2 V'(x)-\frac nx\ri)^2 - 2 \res{x=\infty}x^{-K-1}  {\rm
Tr}\le(Y^{-1}Y'\, T' \ri)  = \cr
&&=  \frac 1 2  \sum_{J=0}^{K} u_{J+1}u_{K+1-J} -2 n u_{K+2} -
\sum_{J=1}^{d+1}  \res{x=\infty}u_{J}  {\rm
Tr}\le(Y^{-1}Y'\,\sigma_3 x^{J-K-2}  \ri) = \cr
&&=  \frac 1 2 \sum_{J=0}^{K} u_{J+1}u_{K+1-J} -2 n u_{K+2} -
2\!\!\!\!\!\sum_{J=1}^{d-K-1}u_{J+K+2} J\frac \pa{\pa u_J}
\ln(\tau_{IM,n})
\end{eqnarray}	
We therefore have the relations:
\be
\sum_{J=1}^{d-K-1}u_{J+K+2} J\frac \pa{\pa u_J}
\ln(\mathcal Z_n) \equiv \sum_{J=1}^{d-K-1}u_{J+K+2} J\frac \pa{\pa u_J}
\ln(\tau_{IM,n}) \ ,\qquad  K=0,\dots d-2.\label{viras}
\ee
The vector fields appearing in eq. (\ref{viras}) are
just representations of the negative Virasoro generators
 on the space of functions of the deformation parameters $\{u_K\}$.
\be
\mathbb V_{-K}  =\sum_{J=1}^{d+1-K}
u_{J+K} J\frac \pa{\pa u_J}\ .
\ee
 Eq. (\ref{viras}) can therefore be rewritten as 
\be
\mathbb V_{-K} \ln\le(\mathcal Z_n(V)\ri) = \mathbb V_{-K}
\ln\le(\mathcal \tau_{IM,n}(V)\ri) \ ,\ \ 2\leq K\leq \dots d\ .
\ee
Since  the vector fields  $\mathbb V_{0},...,\mathbb V_{-d}$ are
linearly independent
\be
\mathbb V_{0}\wedge ... \wedge \mathbb V_{-d} =(d+1)! \le(u_{d+1}\ri)^{d+1}
\pa_{u_1}\wedge\dots\wedge \pa_{u_{d+1}}\ , 
\ee
and $\mathcal Z_n(V)$, $\tau_{IM,n}$ depend on the $d+1$ parameters $u_1,\dots
,u_{d+1}$, we may relate them up to a multiplicative constant if 
 we determine the effect of the two remaining operators.
Now $\mathbb V_{0}$ and $\mathbb V_{-1}$ generate
dilations and translations respectively
\be
\mathbb V_{0} V(x) =x\pa_x
V\le(x\ri) \ ,\ \mathbb V_{-1} V(x) =\pa_x V(x)\ .
\ee
Moreover it is clear from its definition (\eq{partit}) that the partition function must satisfy
\be
\mathbb V_{-1} \mathcal Z_n = 0\ ,\ \ \mathbb V_{0} \mathcal Z_n
=-n^2 \mathcal Z_n \ .
\ee
On the other hand the definition (\ref{tauisom}) of the
isomonodromic tau function implies that it is invariant under translations and
dilations of $x$ 
so that
\be
\mathbb V_{0} \tau_{IM,n} =\mathbb V_{-1} \tau_{IM,n}= 0\ .
\ee
It follows that 
\be
\ln\le({\mathcal Z_n}\ri) = \ln({\tau_{IM,n}})- \frac
{n^2}{d+1}\ln(u_{d+1}) + {\rm const},
\ee
which gives the proof of the theorem up to a constant multiplicative
factor. Q.E.D. 

\vskip 3pt
\br Thm. \ref{mainthm} was proved 
in the special case in which $V(x) = g_{1}x^2 + g_{2} x^4$ 
in \cite{isomIts}.
\er
\br
For any fixed $n$, the set of $\varkappa$ for which orthogonal polynomials exist up to
degree $n$ is an open set because it is the complement of the
vanishing locus of $n$ homogeneous polynomials
\be
\Delta_m(\varkappa,V)\neq 0\ ,\ \ \forall m\leq n.
\ee
Thus we could also study the deformations with respect to 
parameters $\varkappa$. 
However, these are  not isomonodromic deformations because they change the
data of the Riemann--Hilbert 
 problem that is naturally associated to eq. (\ref{ODE}). A similar
approach in the case of even potentials can be found in \cite{stringIts}.
\er

\appendix
 \renewcommand{\theequation}{\Alph{section}-\arabic{equation}}
\setcounter{equation}{0}
\section{Appendix}
\indent In this appendix we give proofs of Theorems \ref{thDeriv}
and \ref{thdeform}. 
First, we  consider the deformation equations, which  determine the
variation of the orthogonal 
polynomials when the potential $V(x)$ is varied. 

Let us first expand the variation of $\psi_n(x)$ with respect to $u_K$
in the basis  $\left\{ \psi_{n}\right\}$: 
\be\label{dpsidukinfty}
{\pa \over \pa u_K} \psi_n(x) = \sum_{m=0}^\infty \left(U_K\right)_{nm} \psi_m(x) 
\ .
\ee
The semi-infinite matrix $U_K$ is antisymmetric (as seen from 
differentiating \eq{ortho}):
\be\label{Ukantisym}
(U_K)_{nm} = - (U_K)_{mn} \ .
\ee
Taking the derivative of \eq{defpsin} with respect to
$u_K$ and using $\pa p_n/\pa u_K = O(x^{n-1})$ and \eq{Ukantisym}) we obtain
\be\label{Ukdef}
U_K = -{1\over 2K} \left( \left(Q^K\right)_+ - \left(Q^K\right)_- \right)  \ .
\ee
In particular the semi-infinite matrix $U_K$ is of finite band size:
 \be 
(U_K)_{nm} = 0\qquad  \hbox{if}\ \  |n-m|>K\ ,
\ee
 and therefore \eq{dpsidukinfty}
contains only a finite sum.
The diagonal part yields (from \eq{defpsin}):
\be
{1\over K} \left(Q^K\right)_{nn} = {\pa \ln{h_n} \over \pa u_K}\qquad 
 \Rightarrow\qquad  \frac \pa{\pa u_K} \ln \mathcal Z_n = \frac 1 K
\sum_{i=0}^{n-1} \left(Q^K\right)_{ii}\label{traceUk} \ .
\ee
The semi-infinite matrix 
$U_K$ can be used to form a differential system in the same way as for 
 the $\pa_x$ equations, by  expressing any $\psi_{m}$ in terms of $\psi_n$ and
$\psi_{n-1}$ by means of \eq{multx}. We then obtain an equation
\be
{\pa \over \pa u_K} \pmatrix{\psi_{n-1}(x) \cr \psi_n(x)} = {\cal
U}_{K,n}(x) \pmatrix{\psi_{n-1}(x) \cr \psi_n(x)} \ ,
\ee
where ${\cal U}_{K,n}(x)$ is a $2\times 2$ matrix with polynomial
coefficients of degree $\leq K$.
In particular it is easy to compute ${\cal U}_{1,n}(x)$:
\be
{\cal U}_{1,n} (x) = -{1\over 2} \pmatrix{ \beta_{n-1}-x & 2\gamma_n
\cr -2\gamma_n & x-\beta_n} \ .\label{induc}
\ee

The general form of the matrices $\mathcal U_{K,n}$ is given in
Thm. \ref{thdeform} which we now prove.\\
{\noindent \bf Proof of Thm. \ref{thdeform}}
For any semi-infinite matrix $A$, let $A_l$ denote the
$l^{\rm th}$ diagonal above (or below, if $l<0$) the main
diagonal. Recall that $A_+$ (resp. $A_-$) denotes the strictly upper
(resp. strictly lower) triangular part of $A$. 

\medskip

Since the matrix $Q$ has only three non-vanishing diagonals, we have:
\begin{eqnarray}
\left(Q^{K+1}\right)_+ & = & ( Q^K Q)_+ = ( Q^K Q_1)_+ + ( Q^K Q_0)_+
+ ( Q^K Q_{-1})_+  \cr
& = & \left(\left(Q^K\right)_+ + \left(Q^K\right)_0\right) Q_1  +
\left(Q^K\right)_+ Q_0 + \left(\left(Q^K\right)_+ -
\left(Q^K\right)_1\right) Q_{-1}  \cr
& = & \left(Q^K\right)_+ Q +  \left(Q^K\right)_0 Q_1 -
\left(Q^K\right)_1 Q_{-1} 
 \ .
\end{eqnarray}
This, combined with a similar calculation for $\left(Q^{K+1}\right)_-$ implies:
\be
Q^{K+1}_+ - Q^{K+1}_- =  (Q^K_+ -  Q^K_-)Q +  (Q^K_0+Q^K_{-1})Q_1  -
(Q^K_{1} + Q^K_0) Q_{-1}  \ ,
\ee
which, componentwise, reads (using \eq{Ukdef}):
\begin{eqnarray}
-2(K+1) {\pa \over \pa u_{K+1}} \psi_n(x) & = &  -2K {\pa \over \pa
u_{K}} x\psi_n(x)  + \left( Q^K\right)_{nn}( \gamma_{n+1}\psi_{n+1}
-\gamma_n\psi_{n-1}) \cr 
& & + \left( \gamma_n\left( Q^K\right)_{n,n-1}  - \gamma_{n+1}  \left
( Q^K\right)_{n,n+1} \right) \psi_n  \ . 
\end{eqnarray}
Using \eq{multx}:
\bea
-2(K+1) {\pa \over \pa u_{K+1}} \psi_n(x) & = & -2K {\pa \over \pa u_{K}} x\psi_n(x)  
 + \left( Q^K\right)_{nn}( (x-\beta_n)  \psi_{n} -  2\gamma_n \psi_{n-1}) \cr
& & + \left(\gamma_n \left( Q^K\right)_{n,n-1}  - \gamma_{n+1}  \left( Q^K\right)_{n,n+1} \right) \psi_n \\
-2(K+1) {\pa \over \pa u_{K+1}} \psi_{n-1}(x) & = & -2K {\pa \over \pa
u_{K}} x\psi_{n-1}(x) 
+ \left( Q^K\right)_{n-1,n-1}( 2\gamma_{n}  \psi_{n} -  (x-\beta_{n-1}) \psi_{n-1}) \cr
& & + \left(\gamma_{n-1} \left( Q^K\right)_{n-1,n-2}  - \gamma_{n}
\left( Q^K\right)_{n,n-1} \right) \psi_{n-1} \ ,
\eea
which can be further simplified using
\bea
Q^{K+1}_{nn} & = & \gamma_{n+1} Q^K_{n,n+1} + \beta_n Q^K_{nn} + \gamma_n Q^K_{n,n-1} \\
Q^{K+1}_{n-1,n-1} & = & \gamma_{n-1} Q^K_{n-1,n-2} + \beta_{n-1} Q^K_{n-1,n-1} + \gamma_n Q^K_{n,n-1} \ .
\eea
We thus get a recursion formula for ${\cal U}_{K,n}(x)$:
$$
-2(K+1){\cal U}_{K+1,n}(x)  + 2K x {\cal U}_{K,n}(x)  = 
$$
\be
\pmatrix{- \left(xQ^K-Q^{K+1}\right)_{n-1,n-1} & 0 \cr
  0 & \left(xQ^K - Q^{K+1}\right)_{nn} }
+ 2\gamma_n
\pmatrix{ - Q^K_{n-1,n}  & Q^K_{n-1,n-1} \cr -Q^K_{nn} & Q^K_{n,n-1} } \ ,
\ee
from which the result follows by induction on $K$ and using
\eq{induc}. Q.E.D.

 \vskip 3pt
As a corollary to Thm. \ref{thdeform} we can complete the proof of Thm. \ref{thDeriv}.\\
{\bf Proof of Thm. \ref{thDeriv}.}
Since $V(x)$ is a polynomial of degree $d+1$,  eq. (\ref{PV}) implies that:
\be
P = \sum_{K=0}^{d} u_{K+1}\, K \,U_K \ ,
\ee
and therefore
\be
\DD_n(x) =  \sum_{K=1}^{d} u_{K+1}\, K \,{\cal U}_{K,n}(x) \ .
\ee
Using Thm. \ref{thdeform} and \eq{eqmvt}, one immediately finds the
expression (\ref{formulafordd}) for $\DD_n(x)$. Q.E.D.
\section{Appendix}
\label{B}
 \setcounter{equation}{0}
We give here a simple proof of the fact that the partition function $\mathcal
Z_n(V)$ is a KP tau function in the sense of \cite{sato, segalwilson};
that is, the determinant of a projection operator acting on elements
of a suitably defined Grassmannian, evaluated along the orbits of an
Abelian group. The interest in rederiving this well-known result is
that the method used, which just involves a simple formal matrix
calculation, naturally leads to a new interpretation of the partition
function as the Fredholm determinant of a simple integral operator.

Let 
\be
\mathcal H = \mathcal H_n +\mathcal H_+\ ,
\ee
where $\mathcal H_n = {\rm span}\{z^j\}_{0\leq j\leq n-1}$ and
$\mathcal H_+$ is a suitably defined completion (whose details are
irrelevant for this sort of formal computation) of the space ${\rm
span}\{z^j\}_{j\geq n}$, with $z\in \C$ viewed as a variable defined
along our cycle $\varkappa \Gamma$.
(Note that we could have included the negative powers
$\{z^j\}_{j<0}$ in the definition of $\mathcal H$, but for the case at
hand, these contribute nothing to the computation of the tau-function
and it is simplest to suppress them from the start.)
 Denote by $\widetilde w_0$ the
$Gl(\mathcal H)$ element defined by 
\be
\begin{array}{lccl}
\widetilde w_0:&\mathcal H &\longrightarrow& \mathcal H\\
\widetilde w_0:&\sum_{j\geq 0} a_j z^j & \mapsto &  \sum_{j\geq 0} a_j p_j(z)\ , 
\end{array}
\ee
where $\{p_j(z)\}$ are the monic orthogonal polynomials associated to the moment
functional $\mathcal L_{\varkappa\Gamma}$ for an initial value of the
parameters $\{u_k\}$.
Within the natural basis $\{ z^j\}_{j\in \N}$ the group element
$\widetilde w_0$ is represented by the upper triangular matrix 
\be
\widetilde W_0 := 
\mathop{\le[\begin{array}{c|c}
W_n & W_{n+}\cr\hline
0 & W_+
\end{array} \ri]}^{\displaystyle{n\ \ | \ \ +}}\begin{array}{c}
n\\ \hline  + \end{array}
\ee
whose $j$-th column consists of the expansion coefficients in 
\be
p_j(z) := \sum_{k=0}^{j} W_{kj} z^k\ .
\ee
(Note that the blocks $W_n,W_{+}$ are upper triangular with diagonal
elements all equal to $1$.)

  We consider the Grassmannian $Gr_+(\mathcal H)$ of subspaces of $\mathcal H$
that are ``commensurable'' with $\mathcal H_+\subset \mathcal H$ in the sense
of \cite{segalwilson}, and the commuting flows induced on it by the
action of the Abelian group 
\be
\Gamma^+:=\le\{\widetilde \gamma({\bf t}) = \exp\le(\sum_{J\geq 1} \frac 1J t_J
z^J\ri)\ri\}\,\subset\, Gl(\mathcal H)\ ,
\ee
(where ${\bf  t} = (t_1,t_2,\dots)$),
acting by multiplication:
\be
\begin{array}{ccc}
\Gamma^+\times \mathcal H &\longrightarrow & \mathcal H\\
(\widetilde \gamma({\bf t}),f) & \mapsto & \widetilde\gamma({\bf t}) f\ .
\end{array}
\ee
Within the standard basis, the flows on $\mathcal H$ are represented
by semi--infinite lower triangular matrices 
\be
 \gamma({\bf t} ) := \le[\begin{array}{c|c}
A&0\\ \hline
B&C 
\end{array} \ ,
\ri]
\ee
where the diagonal blocks are again lower triangular with $1$'s along
the diagonal. 
The initial element $w_0\in Gr_+(\mathcal H)$ is taken as 
\be
w_0:= {\rm span}\{p_j(z)\}_{j\geq n}\ .
\ee 
In homogeneous coordinates, relative to the standard basis, this is
represented by the $(n+\infty)\times \infty$ rectangular matrix 
\be
W_0:= \widetilde W_0\le[\matrix{0\cr \1_+}\ri]\ .
\ee

The induced flow on $Gr_+(\mathcal H)$ takes $w_0$ to the element
with homogeneous coordinates 
\be
W({\bf  t})=\gamma({\bf t})W_0\ .
\ee 
The corresponding KP tau function $\tau_{W_0}({\bf t})$ is given by the
Pl\"ucker coordinate corresponding to the subspace 
\be
\tau_{W_0}({\bf t}) = \det\le(B\,W_{n+} + C\,W_+\ri)\ .
\ee
Since all diagonal blocks have unit determinants, this equals 
\be
\tau_{W_0}({\bf t}) = \det\Big( \1_\infty + BW_{n+}{W_+}^{-1}
C^{-1}\Big) 
= \det\Big( \1_n + W_{n+}{W_+}^{-1}C^{-1}B\Big) 	\label{b11a}
\ee
where the second equality, expressing $\tau_{W_0}({\bf t})$ as a finite
$n\times n$ determinant, follows from the Weinstein--Aronszajn
identity.

The partition function $\mathcal Z_n$ in (\ref{partit}) is also expressible as
an $n\times n$ determinant  of the matrix of moments
$\mathfrak M$
\be
\mathfrak M_{ij} := \mathcal L_{\varkappa \Gamma}(x^{i+j})\ ,
\ee
which in our notation may be expressed at the parameter values
\be
u_J = u_J^{(0)} + t_J,\ J=1,\dots,d+1\ ,
\ee
as
\be
\mathfrak M = \mathfrak M_0\,\gamma^{-1}({\bf t})\ ,
\ee
where 
\bea
\mathbf t:= \le(t_1,t_2,\dots,t_{d+1},0,0,\dots\ri)\ ,\\
\mathfrak M_0 = {W_0^t}^{-1} \, H_0\, {W_0}^{-1} 
\eea
is the initial value of $\mathfrak M$ (at ${\bf  t} = { 0}$) and 
\be
H_0:= {\rm diag}(h_0,h_1,\dots) = \le[\matrix{H_{0,n}&0\cr
0&H_{0,+}}\ri]
\ee 
is the initial value of the matrix of normalization factors.

Let 
\be
\widehat\imath_n:\mathcal H_n\hookrightarrow \mathcal H\ ,
\qquad\widehat \pi_n:\mathcal H\to \mathcal H_n\ ,
\ee
 be the natural injection and projection represented by the matrices:
\be
\imath_n:= \le[\matrix{\1_n\cr 0}\ri]\qquad \qquad \pi_n:=
\le[\matrix{\1_n & 0}\ri]\ .
\ee
Then, by \eq{partit}
\bea
&\frac 1{n!}\mathcal Z_n &= \det(\pi_n\mathfrak M\imath_n) = \det\le( \pi_n
{W_0^t}^{-1} H_0 {W_0}^{-1} \gamma^{-1} \imath_n\ri) \\
 &&=\det(H_{0,n})\det \le(\1_n + W_{n+} {W_+}^{-1} C^{-1}B  \ri) \ ,
\eea
where the second line follows from the fact that 
\be
\det(A) = \det(C) = \det (W_n) = \det (W_+) = 1\ .
\ee
Therefore 
\be
\tau_{W_0}({\bf  t}) = \frac {\mathcal Z_n({\bf t})}{\mathcal Z_n(0)} 
\ . \qquad\qquad {\rm Q.E.D.}
\ee
\indent We may now use this calculation to deduce a representation of
$\mathcal Z_n({\bf t})$ as a Fredholm determinant.
First note that from the definitions of the matrices $A, B,
C,W_{n+},W_+$, we may view these as representations of the operators:
\bea
&&\begin{array}{rccl}
\widehat B: &\mathcal H_n&\longrightarrow& \mathcal H_+\\[5pt]
\widehat B:& \sum_{j= 0}^{n-1} a_j z^j&\mapsto & \sum_{j= 0}^{n-1}
 a_j \big(\widetilde \gamma({\bf  t})z^j\big)_{\geq n}
\end{array}\ \\[5pt]
&& \begin{array}{rccl}
\widehat C: &\mathcal H_+&\longrightarrow& \mathcal H_+\\[5pt]
\widehat C: &\sum_{j\geq n} a_j z^j&\mapsto & \sum_{j\geq n} a_j \widetilde\gamma({\bf  t})z^j
\end{array}\ \\[5pt]
&& \begin{array}{rccl}
\widehat W_+:& \mathcal H_+&\longrightarrow& \mathcal H_+\\[5pt]
\widehat W_+:& \sum_{j\geq n}a_j z^j&\mapsto & \sum_{j\geq n}a_j\big(p_j(z) \big)_{\geq n}
\end{array}\ \\[5pt]
&& \begin{array}{rccl}
\widehat W_{n+}: &\mathcal H_+&\longrightarrow& \mathcal H_n\\[5pt]
\widehat W_{n+}: &\sum_{j\geq n}a_j z^j&\mapsto & \sum_{j\geq
n}a_j\big(p_j(z) \big)_{<n}\ ,
\end{array}
\eea
where $(f(z))_{\geq n}$ and $(f(z))_{<n}$ denote the parts of the
Taylor series expansion of $f\in \mathcal H$ lying in $\mathcal H_+$
and $\mathcal H_n$ respectively.
It follows that the operator :
\be
\widehat K: \mathcal H_+\to\mathcal H_+\ 
\ee
represented above by the matrix appearing in eq. (\ref{b11a}):
\be
-BW_{n+}{W_+}^{-1}C^{-1}
\ee
is just the rank-$n$ Fredholm integral operator
\be
\widehat K\, f(z) = \int_{\varkappa \Gamma}\!\!\!{\rm d}w\, {\rm
e}^{-\sum_{k=1}^{d+1} \frac{u_K^{(0)}}{K} w^K}\,K(z,w)f(w)\
,\qquad f\in \mathcal H_+
\ee
with kernel
\be
K(z,w) = \sum_{j=0}^{n-1}\frac 1{h_j} \le({\rm e}^{\sum_{K=1}^{d+1}t_K z^K/K}
p_j(z) \ri)_{\geq n} p_j(w){\rm e}^{-\sum_{K=1}^{d+1}t_K w^K/K}\ ,
\ee
and hence, by \eq{b11a}, 
\be
\tau_{W_0}({\bf t}) = \det\le(\1 - \widehat K\ri)\ .
\ee\vskip 15pt
\noindent{\bf \large Acknowledgments} \\
The authors would like to thank A. Its and A. Orlov for helpful remarks and for 
drawing attention to related earlier works.


\begin{thebibliography}{99}

\bibitem{ASvM}
M. Adler, P. van Moerbeke, T. Shiota,  ``Random matrices, Virasoro algebras
and noncommutative KP'', {\em Duke Math. J.} {\bf 94}  379--431 (1998). 

\bibitem{AvM}
M. Adler, P. van Moerbeke, ``Hermitian, symmetric and symplectic
random ensembles: PDEs for the distribution of the spectrum'', {\em
Ann. of Math.} (2) {\bf 153}  no. 1, 149--189 (2001). 

\bibitem{aratyn} H.~Aratyn and J.~van de Leur, ``Integrable Structure
behind WDVV Equations'', contribution to NEEDS 2001 Conference, hep-th/0111243,
to appear in {\em Theor. Math. Phys}.

\bibitem{bauldry} W. Bauldry, ``Estimates of asymmetric Freud
Polynomials on the real line'', {\em J. Approx. Theory} 
{\bf 63}, 225-237 (1990).

\bibitem{BEH} M. Bertola, B. Eynard and J. Harnad, ``Duality,
Biorthogonal Polynomials and Multi--Matrix Models'',
{\em Comm. Math. Phys.} (in press, 2002), CRM-2749 (2001), 
Saclay-T01/047, nlin.SI/0108048.

\bibitem {BlIt} P. Bleher, A. Its, ``Semiclassical asymptotics of
orthogonal polynomials, Riemann-Hilbert problem, 
and universality in the matrix model'' {\em Ann. Math.}
 (2) {\bf 150}, no. 1, 185--266 (1999). 

\bibitem {BlIt1} P. Bleher, A. Its, 
``On asymptotic analysis of orthogonal polynomials via the Riemann-Hilbert
method'', Symmetries and integrability of difference equations
(Canterbury, 1996), 165--177, {\em London Math. Soc. Lecture Note Ser.},
{\bf 255}, Cambridge Univ. Press, Cambridge, 1999.

\bibitem{bonan} S.S. Bonan, D.S. Clark, ``Estimates of the Hermite and
the Freud polynomials'', {\em  J. Approx. Theory} {\bf 63}, 210-224 (1990).

\bibitem{BD} A. Borodin, P. Deift, ``Fredholm determinants,
Jimbo-Miwa-Ueno Tau-functions, and representation Theory'',
{\em Comm. Pure Appl. Math.} {\bf 55}  no. 9 1160--1230 (2002).  

\bibitem{ZJDFG} P. Di Francesco, P. Ginsparg, J. Zinn-Justin, ``2D
Gravity and Random Matrices'' {\em Phys. Rep.} {\bf 254}, 1 (1995). 

\bibitem{doug} M. R. Douglas, S. H.  Shenker, ``Strings in less than one 
dimension'', {\em Nuclear Phys. B} {\bf 335}, no. 3, 635--654 (1990). 

\bibitem{doug1} M. R. Douglas, ``Strings in less than one dimension and the 
generalized KdV hierarchies'',  {\em Phys. Lett. B} {\bf 238}, no. 2-4, 176--180
(1990).

\bibitem{courseynard} B. Eynard ``An introduction to random matrices'', 
lectures given at Saclay, October 2000, notes available at 
http://www-spht.cea.fr/articles/t01/014/.

\bibitem{FIK} A. Fokas, A. Its, A. Kitaev, ``The isomonodromy approach
to matrix models in 2D quantum gravity'', {\em Commun. Math. Phys.}
{\bf 147}, 395--430 (1992). 

\bibitem {orlov} A. Gerasimov, A. Marshakov,A.  Mironov,
A.  Morozov, A. Orlov, ``Matrix models of two-dimensional gravity and Toda 
theory'', {\em Nucl. Phys. B} {\bf 357}, no. 2-3, 565--618 (1991).

\bibitem{H1} J. Harnad, "On the Bilinear Equations for Fredholm Determinants
Appearing in Random Matrices", {\em J. Nonlinear Math. Phys.} {\bf  9}, (In
press, 2002).

\bibitem{HI} J. Harnad and A. R. Its  ``Integrable Fredholm Operators
and Dual Isomonodromic  Deformations'', {\it Commun. Math. Phys.} {\bf 226},
497-530 (2002). 

\bibitem{HTW} J.  Harnad,  C.~A. Tracy,  and  H. Widom, 
``Hamiltonian Structure of Equations Appearing in Random Matrices'' , in:
{\em  Low Dimensional Topology and Quantum Field Theory}, ed. H. Osborn,
(Plenum,  New York  1993), pp. 231--245.

\bibitem{ismail} M.E.H. Ismail, Y. Chen, ``Ladder operators and
differential equations for orthogonal polynomials'', {\em J. Phys. A.} {\bf
30}, 7818-7829 (1997).

\bibitem{isomIts} A. R. Its, A. V.  Kitaev, A. S. Fokas  ``Matrix models of 
two-dimensional quantum gravity and isomonodromy solutions of
    ``discrete Painleve equations'' '',{\em  Zap. Nauch. Sem. LOMI},  
{\bf 187}, 3-30 (1991) (Russian), translation in {\em  J. Math. Sci.} {\bf 73},
no. 4, 415--429  (1995).

\bibitem{stringIts}  A.R. Its, A.V. Kitaev, and A.S. Fokas,  ``An 
isomonodromic Approach in the
Theory of Two-Dimensional Quantum Gravity'', {\em Usp. Matem. Nauk}, {\bf 45}, 
6 (276), 135-136 (1990), (Russian), translation in {\em Russian Math.
Surveys}, {\bf 45}, no. 6, 155-157 (1990).

\bibitem{JMU} M. Jimbo, T. Miwa and K. Ueno, ``Monodromy Preserving 
Deformation of Linear Ordinary Differential Equations with Rational
Coefficients I.'', {\em Physica} {\bf 2D}, 306-352 (1981).

\bibitem{marce} F. Marcell\'an, I. A. Rocha, ``Complex Path Integral
Representation for Semiclassical Linear Functionals'', {\em J. Appr. Theory}
{\bf 94}, 107--127, (1998) 

\bibitem{marce1} F. Marcell\'an, I. A. Rocha, ``On semiclassical
linear functionals: Integral representations'',
{\em J. Comput. Appl. Math.} {\bf 57}, 239--249 (1995). 

\bibitem{mehta} M. L. Mehta, ``Random Matrices'', 2nd edition
(Academic Press, New York 1991).

\bibitem{moore} G. Moore, ``Geometry of the string equations'',
{\em Comm. Math. Phys.} {\bf 133}, no. 2, 261--304 (1990).

\bibitem{morozov} A. Morozov, 
``Matrix models as integrable systems'',  Particles and fields (Banff,
AB, 1994), 127--210, {\em CRM Ser. Math. Phys.}, Springer, New York, 1999.

\bibitem{palmer} J. Palmer, 
``Deformation analysis of matrix models'',
{\em Physica D} {\bf 78} 166--185 (1994).

\bibitem{sato} M. Sato, Y. Sato,
``Soliton equations as dynamical systems on infinite-dimensional
Grassmann manifold'',  Nonlinear partial differential
equations in applied science (Tokyo, 1982), 259--271,
 {\em North-Holland Math. Stud.} {\bf 81}, North-Holland, Amsterdam, 1983. 

\bibitem{segalwilson} G. Segal, G.  Wilson,  
``Loop groups and equations of KdV type'', 
{\em Inst. Hautes \'Etudes Sci. Publ. Math.} {\bf 61}, 5--65 (1985).

\bibitem{Szego} G. Szeg\"o, ``Orthogonal Polynomials'', AMS, Providence, 
Rhode Island, (1939).

\bibitem{TW1} C.~A. Tracy,  H. Widom, ``Introduction to Random
Matrices'',  in: {\em 	Geometric and Quantum Methods in Integrable Systems.}
 Springer Lecture Notes in Physics {\bf	424}, pp. 103--130
(ed. G. F. Helminck, Springer--Verlag, N.Y.,  Heidelberg,  1993).

\bibitem{TW2} C.~A. Tracy,  H. Widom, ``Level Spacing Distributions and
the Airy Kernel'', {\it Commun.~Math.~Phys.\/} {\bf 159}, 151--174
(1994); ``Level Spacing Distributions and the Bessel Kernel'', {\em
ibod./} {\bf 161}, 289--309 (1994).

\bibitem{vM} P. van Moerbeke, ``The spectrum of random matrices and integrable
systems'', Group21, Physical applications and Mathematical aspects of
Geometry, Groups and Algebras, Vol II, 835--852, eds., Doebner,
 Scherer, Schulte (World Scientific, Singapore, 1997).
\end{thebibliography}
\end{document}